\documentclass[%
 reprint,
superscriptaddress,
 amsmath,amssymb,
 aps,
prl,
]{revtex4-1}

\usepackage{graphicx}
\usepackage{dcolumn}
\usepackage{bm}
\usepackage{natbib}
\usepackage{epstopdf} 
\usepackage{enumitem}

\begin{document}

\preprint{APS/123-QED}

\title{Self-referenced coherent diffraction x-ray movie of \AA ngstrom- and femtosecond-scale atomic motion}
\author{J. M. Glownia} \thanks{J.M. Glownia and A. Natan contributed equally to this work.} %
\affiliation{Linac Coherent Light Source, SLAC National Accelerator Laboratory, Menlo Park, CA 94025}
\affiliation{Stanford PULSE Institute, SLAC National Accelerator Laboratory Menlo Park, CA 94025}
\author{A. Natan} \thanks{J.M. Glownia and A. Natan contributed equally to this work.}%
\author{J. P. Cryan}%
\author{R. Hartsock}%
\affiliation{Stanford PULSE Institute, SLAC National Accelerator Laboratory Menlo Park, CA 94025}
\author{M. Kozina}
\author{M. P. Minitti}%
\author{S. Nelson}%
\author{J.~Robinson}%
\author{T. Sato}%
\author{T. van Driel}
\author{G. Welch}%
\affiliation{Linac Coherent Light Source, SLAC National Accelerator Laboratory, Menlo Park, CA 94025}
\author{C. Weninger}%
\affiliation{Stanford PULSE Institute, SLAC National Accelerator Laboratory Menlo Park, CA 94025}%
\affiliation{Linac Coherent Light Source, SLAC National Accelerator Laboratory, Menlo Park, CA 94025}
\author{D. Zhu}%
\affiliation{Linac Coherent Light Source, SLAC National Accelerator Laboratory, Menlo Park, CA 94025}
\author{P. H. Bucksbaum}
\affiliation{Stanford PULSE Institute, SLAC National Accelerator Laboratory Menlo Park, CA 94025}%
\affiliation{Departments of Physics, Applied Physics, and Photon Science, Stanford University, Stanford, CA 94305.}




\date{\today}

\begin{abstract}
Time-resolved femtosecond x-ray diffraction patterns from laser-excited molecular iodine are used to create a movie of intramolecular motion with a temporal and spatial resolution of
$30~$fs and $0.3$ \AA .
This high fidelity is due to interference between the
non-stationary 
excitation and the
stationary 
initial charge distribution.
The initial state is used as the local oscillator for heterodyne amplification of the excited charge distribution to retrieve real-space movies of atomic motion on \AA ngstrom  and femtosecond scales.
This x-ray interference has not been employed to image internal motion in molecules before.
Coherent vibrational motion and dispersion, dissociation, and rotational dephasing  are all clearly visible in the data, thereby demonstrating the stunning sensitivity of heterodyne methods.
\end{abstract}

\pacs{Valid PACS appear here}
\maketitle



High brightness ultrafast hard x-ray free electron lasers (FELs) can perform time-resolved x-ray diffractive imaging.
Recent demonstrations of time-resolved crystal diffraction or time-resolved non-periodic imaging illustrate the power of these sources to track  \AA ngstrom-scale motion~\cite{trigo_fourier-transform_2013, kupitz_serial_2014}.
These have spurred new insights in broad areas of science, but have not
fully realized the potential of x-ray FELs to image molecules with simultaneous sub-\AA ngstrom and few-femtosecond resolution.
Previous x-ray or electron scattering experiments have used correlations between simulations and data to extract femtosecond molecular dynamics information ~\cite{
minitti_imaging_2015,kupper_x-ray_2014, Yang_Guehr_2016, Blaga_Nat_2012, Boll_PRA_2013}.

Here we propose and demonstrate an imaging method that employs a universal but unappreciated feature of time-resolved hard x-ray scattering that dramatically improves reconstructed images of charge motion, and enables femtosecond and sub-\AA ngstrom x-ray movies.
The method relies on the ``pump-probe'' protocol, where motion is initiated by a short ``start'' pulse, and then interrogated at a later time by a ``probe'' pulse.  The pumped fraction is small,
and the unexcited fraction is our heterodyne reference \cite{Wu_Liu_Rose-Petruck_Diebold_2012}.

When a gas of $N$ identical molecules in a thermal distribution is excited with probability $a$ from the ground state $g$ to an excited state $e$, only a fraction $aN$ molecules are in $e$ but there is no information about which ones.  If we scatter x-rays from this system, the elastic scattering amplitude~\cite{Dixit_Vendrell_Santra_2012}
\begin{equation}\label{f(Q,t)}
f(\vec{Q},t)=\int d^{3}x\rho(\vec{x},t)e^{i \vec{Q} \cdot \vec{x}}, 
\end{equation}
is the normalized sum of $f^{(g)}$ or $f^{(e)}$ from all $N$ molecules in all $M$ possible excitation configurations.
Here $\rho$~
is the instantaneous charge density,
$\vec{Q}$ is the photon momentum transfer, and
$M=\binom{N}{aN}=N!/[(N-aN)!aN!]$.
This sum can be expanded:
\begin{equation}\label{doublesum}
f(\vec{Q})= (1/M)\sum_{i=1}^{M}
\left[\sum_{j=1}^{aN}f^{(e)}_{ij}(\vec{Q}) +
\sum_{j=1}^{(1-a)N}f^{(g)}_{ij}(\vec{Q})\right].
\end{equation}
The order of summing can be re-arranged so that the factor $1/M$ cancels the sum over $i=1,M$, leaving:
\begin{equation}\label{singlesum}
f(\vec{Q})=\sum_{j=1}^{N}\left[af^{(e)}_{j}(\vec{Q})+(1-a)f^{(g)}_{j}(\vec{Q})\right]
\end{equation}
The square of this amplitude is the intensity on the detector.  Cross terms between different molecules average out due to their random position in the gas, so the scattered intensity $I$ is linear in the number of molecules N:
\begin{equation}\label{S(Q)}
I(\vec{Q}) = |f(\vec{Q})|^2=N|af^{(e)}(\vec{Q})+(1-a)f^{(g)}(\vec{Q})|^2
\end{equation}
This signal is an incoherent sum of the coherent diffraction from each molecule.
Such a system is described by a quantum density matrix for coherent rovibrational excitation with incoherent mixtures of ground and excited electronic states.
Eq.\ref{S(Q)} differs from the result for an inhomogeneous gas mixture where there are no intramolecular cross terms and the intensity distributions of the two species simply add.

\begin{figure}
  \centering
  \includegraphics[width=3.5in]{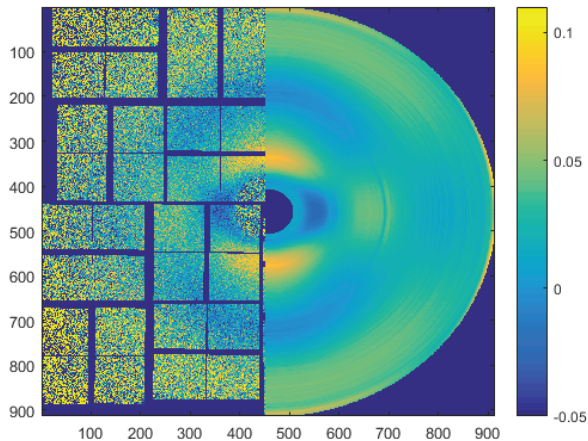}
  \caption{Left: Half of the LCLS 2.5 megapixel array detector (CSPAD \cite{herrmann2013cspad}) showing
the fractional deviation from the mean scattering signal recorded in each pixel at a pump-probe delay of
150~fs, integrated over 100 x-ray pulses.
Right: The Legendre polynomial fit obtained by applying  Eq.~\ref{legendre} to the data at this time delay. The scattering vs. time delay appears as a movie in Supplemental Material \cite{supplemental}.
}
\label{inset}
\end{figure}

The key insight in Eq.\ref{S(Q)} is that scattering from the excited fraction in each molecule interferes with scattering from its initial state fraction,
producing holographic fringes.
The scattering from the excitation alone without ground-state interference goes as $a^2 N$ according to Eq.
\ref{S(Q)};
but the modulation due to holographic interference has a peak-to-peak amplitude proportional to $4aN$.  This  increase factor of  $4/a$ in the pattern of x-rays on the detector makes it possible to create high fidelity images of the excited charge distribution using heterodyne deconvolution to extract the signal.

Formal descriptions of time-resolved x-ray diffraction in small molecules have not discussed the importance of this self-referenced interference~\cite{Ben-Nun_Cao_Wilson_1997, Cao_Wilson_JPCA_1998, Bratos_Vuilleumier_JCP_2002, Henriksen_Moller_JPCB_2008, Lorenz_Moller_Henriksen_2010, Debnarova_Techert_Schmatz_JCP_2010, Dixit_Vendrell_Santra_2012, Ben-Nun_Martínez_Weber_Wilson_1996}.
Eq.~\ref{S(Q)} has been noted previously, but has not been implemented for molecular movies~\cite{Ben-Nun_Martínez_Weber_Wilson_1996, Woerner_Zamponi_Ansari_Dreyer_Freyer_Prémont-Schwarz_Elsaesser_2010,Vrakking_Elsaesser_2012,Reis_Lindenberg_2006}.
The initial reference distribution is extracted from negative delay data, when the probe sees the initial distribution.
The deconvolved signal is a \textit{de-novo} molecular movie.

A demonstration of coherent self-referenced time-resolved imaging was performed at the X-ray Pump Probe~(XPP) facility at the Linac Coherent Light Source~(LCLS) \cite{Chollet}. Molecular iodine vapor was resonantly excited with a short laser pulse from the
X$(^1\Sigma_{g}^{+})$ state to the B$(^3\Pi_{0u}^{+})$ state \cite{Tellinghuisen_1973}. This excites a coherent vibrational wavepacket \cite{Krause_Whitnell_Wilson_Yan_Mukamel_1993}.

\begin{figure}
  \centering
  \includegraphics[width=3.5in]{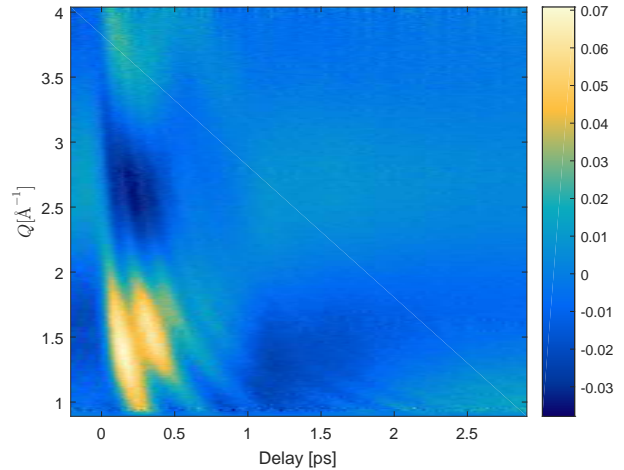}
  \caption{$\beta_{2}(Q,t)$ as defined in Eq.~\ref{legendre}. This term captures most of the excited state scattering signal. The time-averaged value is subtracted, and the scale is the fraction of modulation due to the holographic interference between the excited state and the reference ground state.  The principal features are long-period oscillations in Q that are due to the B state, and much shorter period oscillations in Q caused by dissociation.
  }\label{beta2(Q,t)}
\end{figure}

We apply a standard correction to remove the
effect of the angle dependence of the Thomson scattering cross section in the horizontal scattering plane  
due to the LCLS linearly polarized x rays, and we rebin in $(Q,\theta)$ coordinates.
Data from each radial value are fit to a Legendre polynomial basis (Fig.\ref{inset})
\begin{equation}\label{legendre}
I(Q,\theta,t) = A(Q,t)\left[1+\sum_{n=1}^3 \beta_{2n}(Q,t) P_{2n}(\cos(\theta))\right]
\end{equation}

The apparatus for gas phase scattering has been described previously~\cite{Budarz_Minitti_Cofer-Shabica_Stankus_Kirrander_Hastings_Weber_2016}. The pump pulse~
($520\pm5$~nm,
20 $\mu$J,
120~Hz,
$50$~fs,
vertical polarization,
$100\mu$m beam diameter)
was created
by an optical parametric amplifier.
The probe pulse
($9.0$~keV,
$2$~mJ,
$120$~Hz,
$40$~fs,
horizontal polarization,
$30 \mu m$ beam diameter)
was a spatially coherent beam of x-rays provided by the LCLS.
The co-propagating cross-polarized beams were focused into a windowless iodine cell inside a larger vacuum enclosure with a sapphire/beryllium output window.
The perpendicular beam polarizations ensure that the $B$-state modulation is in a direction where the  x-ray scattering cross section is insensitive to angle.
The cell was heated to~$100^{\circ}$~C,
with a column density of $\sim 10^{18}$~cm$^{-2}$. The photoexcitation fraction of $\sim$10\% depends on the photon fluence, attenuation length, beam overlap, and the wavelength-dependent cross section \cite{Saiz-Lopez_Saunders_Joseph_Ashworth_Plane_2004}. The X-ray attenuation was 50\% from transmission losses and 8\% from iodine photoabsorption. Approximately 0.4\% of the remaining x rays undergo iodine elastic scattering, and 2\% of these ($10^{7}$ x rays per pulse) scatter at angles that intercept the 2.3 megapixel silicon array (Cornell-SLAC Pixel Array Detector  \cite{herrmann2013cspad}) detector.
Up to 50 scattered x-rays per pulse per pixel were detected.

\begin{figure*}
  \centering
  \includegraphics[width=6in]{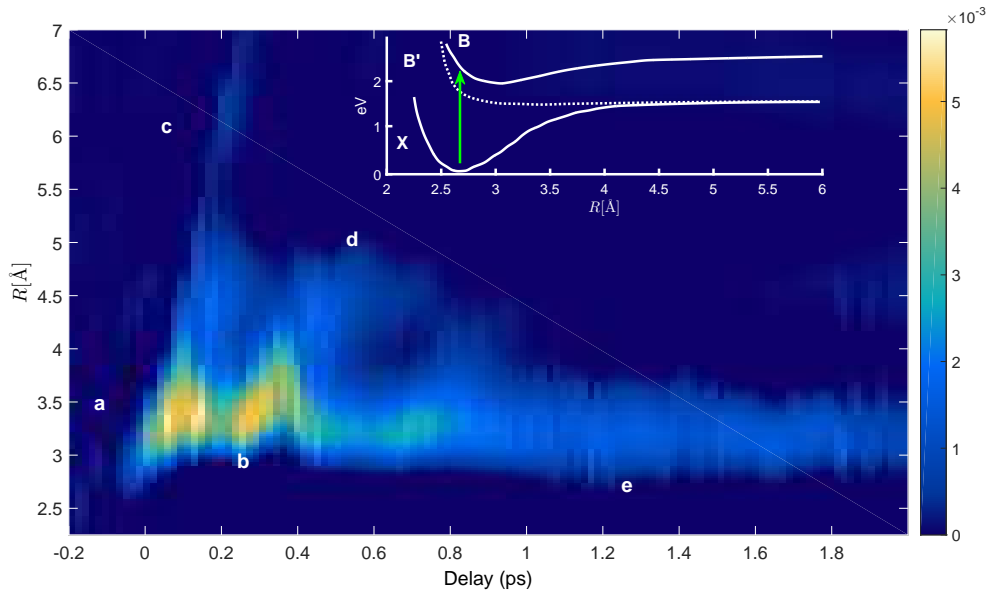}
  \caption{Extracted excited-state charge distribution vs time, for R from 2.3 to 7 \AA~ and time delays out to $2$ ps. This ``movie'' was extracted from $\beta_{2}(Q,t)$ (see Fig. \ref{beta2(Q,t)}) using Eqns.~\ref{f(Q,t)}-\ref{BB(z,t)}.
Bound-state wavepacket oscillations, dissociation, and rotational dephasing are clearly visible. Letters refer to features described in the text. The scale is proportional to the excitation in the $\beta_2$ channel. Inset: Excitation path and iodine molecular potentials.
A video is in Supplemental Material  \cite{supplemental}.
}\label{movie}
\end{figure*}

The x-ray scattering amplitude in Eq.\ref{f(Q,t)}
depends on
the instantaneous charge density
$\rho(\vec{x},t)$~\cite{Dixit_Vendrell_Santra_2012}.
Most of the 53 electrons in iodine are in core orbitals, so the x-rays scatter primarily from the vicinity of the atomic nuclei, and thus the time-dependent charge density will approximately follow the rovibrational motion of the molecule.
Before excitation all of the iodine molecules are in a thermal state in the X manifold.
The laser pulse creates electronically excited rovibrational wavepackets, mostly on the $B$-state.
A typical example of the fractional change in the x-ray diffraction pattern due to laser excitation
is shown in Fig.~\ref{inset}.
The data were discriminated based on x-ray beam parameters (bunch charge, photon energy, pulse energy, and beam position).

Only the even Legendre polynomials are used because the geometry cannot break the up/down symmetry of the molecular ensemble. Contributions for $n > 3$ are negligible.

The radial modulations in Fig.~\ref{inset} are captured in the $\beta_2(Q,t)$ coefficient of Eq.~\ref{legendre}, plotted in Fig.~\ref{beta2(Q,t)}.
This picks out scattering patterns with the symmetry of an electric dipole excitation, and contains nearly all of the time-varying portion of the total scattered signal.
The large-amplitude modulations in Fig.\ref{beta2(Q,t)} are due to holographic interference between the nonstationary charge distribution of the laser-excited wavepacket and the
stationary 
initial charge distribution.
Heterodyne techniques described below allow us to deconvolve the excitation in space and plot it vs. time in Fig.~\ref{movie} as a movie with femtosecond and \AA ngstrom resolution.

For time delays $t<0$ the x-rays scatter from the iodine before the exciting laser arrives in the sample, and therefore the distribution is
stationary 
and contained in the isotropic $A(Q,t)$ portion of  Eq.~\ref{legendre}.

The features in the movie that follow the excitation pulse at t=0 reveal the detailed quantum  evolution of this system. The letters at the beginning of the following paragraphs refer to labeled areas of Fig.~\ref{movie}.

(a)
A region of approximately 100~fs~(about five discrete pump-probe delay points) around $t=0$ shows where the excited state activity begins.
The Franck Condon region, where the B-state is directly over the X-state, is centered around 2.7~\AA~ in iodine. Charge appears across this region moving rapidly towards the center of the B-state potential at approximately 3 \AA~ and then moving beyond towards the outer turning point, approximately 4.5~\AA~for this excitation wavelength.

(b)
The vibrational oscillations in bound states in the molecule can be observed in some detail. See also wavepacket simulations in Supplementary Material \cite{supplemental}.
The excitation is spread over several hundred $\rm{cm^{-1}}$ ($\sim 40$ meV) by thermal broadening of the initial state. The wave packet is high in the anharmonic portion of the B-state potential, and the bound motion in the B-state appears highly dispersed  \cite{Krause_Whitnell_Wilson_Yan_Mukamel_1993, Chen_Fang_Tagliamonti_Gibson_2011}.

(c)
There is a pulse of dissociating charge that starts near $(R,t) = (2.7$~\AA,~$0)$
and moves rapidly away from the bound region with constant velocity and only 4\% of the total excited charge.
The fringes recorded in Fig.~\ref{beta2(Q,t)} are sufficiently fine to show that this feature has little dispersion out to at least 16\AA, well beyond the range included in Fig.~\ref{movie}.
Its velocity is 16~\AA/ps, corresponding to a kinetic energy release of approximately 0.85~eV, consistent with the separation velocity required for the molecule to dissociate into two ground-state atoms for our excitation wavelength.
This prompt dissociation is consistent with transitions to  a family of repulsive \it ungerade \rm states, one of which is shown in the inset in Fig. \ref{movie} \cite{Tellinghuisen_1973,Mulliken_1971}.

(d)
Local moving peaks in the charge density are observed near the outer turning point at time delays of 0.5-0.7 ps.  Similar cusp-like features are predicted
but have not been observed directly before \cite{Chen_Fang_Tagliamonti_Gibson_2011, Krause_Whitnell_Wilson_Yan_Mukamel_1993}. See simulation in Supplemental Material \cite{supplemental}.

(e)
The mean position of the excited  population reaches a minimum value of 3 \AA~ near 1.2 ps. This is consistent with rotational dephasing of the $\cos^{2}\theta $
alignment created in the excited state.
For iodine at $100^{\circ}$~C the initial prolate alignment along $\hat{z}$ evolves to a nearly isotropic distribution at 1.2~ps
\cite{Broege_Coffee_Bucksbaum_2008, Rosca-Pruna_Vrakking_2002, Rosca-Pruna_Vrakking_2001} in agreement with the
data in Fig. \ref{movie}.
Rotational dephasing also affects the total amount of charge vs. time in Fig. \ref{movie}.  The amplitude decreases as population moves from $\cos^2\theta$ to a more isotropic distribution.  Beyond the point of minimum alignment  at 1.2~ps the signal is only about 1/3 the initial strength. See simulation in Supplemental Material \cite{supplemental}.

The method used to ``invert'' this scattering image uses the heterodyne beating that is evident in figure~\ref{beta2(Q,t)}.
Below we describe the step-by-step procedure for obtaining the movie in Fig. \ref{movie}.

The charge density $\rho(\vec{x},t)$ that appears in Eq. \ref{f(Q,t)}
is the expectation value of the charge density operator in the $|\vec{x}>$ basis, which is the trace of the density matrix over the electronic coordinates multiplied by the electron charge.
This can be divided into an initial charge distribution $\rho_{0} (\vec{x})$ and a time-varying distribution $\rho_{e}(\vec{x},t)$ without loss of generality. This agrees with Eq. \ref{S(Q)} for the x-ray intensity $I(\vec{Q},t)=\left|f(\vec{Q},t)\right|^2$.

We approximate $\rho_{0} (\vec{x})$ in the analysis by $\rho(\vec{x},t<0)$, the charge distribution before the laser excitation.
The object of the analysis is to discover $\rho_{e}(\vec{x},t>0)$, and thereby create a molecular movie.
The precise form of the time-independent initial distribution is easily calculated, but we stress here that its most important feature is that it serves as a time-independent reference in the time-varying scattering pattern in a pump-probe experiment.

The process of extracting the excitation from the measured scattering pattern uses heterodyne deconvolution.
The first step is a 2-dimensional inverse Fourier transform of the scattering image.
This cannot recover the charge distribution directly because the scattering is the squared Fourier transform and therefore has no phase information.
However this is an autocorrelation of the charge distribution:
\begin{eqnarray}\label{AC}
  \mathcal{FT}_{2D}^{-1}(f(\vec{Q},t)f^{*}(\vec{Q},t))&=&\mathcal{AC}[\rho(\vec{x},t)] \nonumber  \\
  &\equiv& \rho(\vec{x},t)\otimes \rho(\vec{x},t)
\end{eqnarray}
The right side of Eq. \ref{AC} has contributions from the time-independent reference and the smaller time-dependent wave packet:
\begin{eqnarray}\label{AC2}
  \mathcal{AC}[\rho(\vec{x},t)] &=& \mathcal{AC}[\rho_{0}(\vec{x})] + \mathcal{AC}[\rho_{e}(\vec{x},t)]\nonumber \\
  && +2 \mathcal{CC}[\rho_{0}(\vec{x}),\rho_{e}(\vec{x},t)]
\end{eqnarray}
Here $\mathcal{CC}$ is a convolution integral (i.e. cross-correlation) $  \mathcal{CC}[\rho_{0}(\vec{x}),\rho_{e}(\vec{x},t)] = \rho_0(\vec{x})\otimes \rho_e(\vec{x},t)$.  The first term in Eq.~\ref{AC2} on the right side is obtained from the $t<0$ measurements and can be subtracted.  The second term is second order in the excitation fraction, and may be neglected if the excitation is small.  We then obtain:
\begin{equation}\label{CC}
  2~\mathcal{CC}[\rho_{0}(\vec{x}),\rho_{e}(\vec{x},t)] \simeq \mathcal{AC}[\rho(\vec{x},t)] - \mathcal{AC}[\rho_{0}(\vec{x})]
\end{equation}
The final step to produce a molecular movie uses the convolution theorem once more to extract $\rho_{e}(\vec{x},t)$:
\begin{eqnarray}\label{BB(z,t)}
\rho_{e}(\vec{x},t) &=&\mathcal{FT}_{2D}^{-1}\left[\frac{\mathcal{FT}_{2D}(\mathcal{CC}[\rho_{0}(\vec{x}),\rho_{e}(\vec{x},t)])}{\mathcal{FT}_{2D}[\rho_{0}(\vec{x})]}\right].
\end{eqnarray}
In this step the initial charge distribution $\rho_{0}(\vec{x})$ is approximated as the thermal population of levels of the $X$-state:
\begin{equation}\label{X initial}
  \rho_{0}(\vec{x})\sim\rho_{X}(\vec{x})=\sum_{v=0}^{\infty} \rho_v (\vec{x})e^{-E_{v}/k_{B}T}.
\end{equation}
This is a compact point-spread function for deconvolution in Eq.~\ref{BB(z,t)}.  The image retrieval is thus similar to deblurring in microscopy. We project Eq.~\ref{CC} onto $P_2(\cos\theta)$ and perform a deconvolution (Lucy-Richardson) along $R$, yielding Fig.~\ref{movie}. This deconvolution is robust for several standard algorithms. The resulting resolution is already comparable to the limits in $Q$ imposed by counting statistics and our scattering geometry.

Self-referencing should be applicable to many small molecules in liquid or gas phase. The requirements are: good statistics since the excitation fraction is small;
Sufficient Q-resolution to resolve single bonds; and some knowledge of the initial state charge distribution.   It could be valuable for photo-initiated molecular energy conversion studies such as thymine photoprotection, retinal isomerization, and cyclohexadiene ring openings, providing  molecular movies at the single bond level with relevant time resolution. Pre-alignment methods can improve the measurement fidelity. Since x-ray scattering only detects charge density and motion, it cannot measure electron energies or spins.  Complementary information  comes from time-resolved electron and x-ray spectroscopies and photoelectron imaging~\cite{StolowReview, SuzukiReview,McFarland_natcomm}.

Future higher energy and higher repetition rate x-ray FELs could increase the fidelity and resolution of  molecular movies. Likewise, the method could  be used equally well with sub-femtosecond x-ray pulses, or with enhanced harmonic radiation from FEL undulators.

We wish to acknowledge useful discussions with Ryan Coffee, Markus Guehr, Lucas Zipp, Andreas Kaldun, Jerry Hastings, Kelly Gaffney, Bob Schoenlein, and David Reis in the preparation of this paper.  This research is supported through the Stanford PULSE Institute, SLAC National Accelerator Laboratory by the U.S. Department of Energy, Office of Basic Energy Sciences, Atomic, Molecular, and Optical Science Program.  Use of the Linac Coherent Light Source (LCLS), SLAC National Accelerator Laboratory, is supported by the U.S. Department of Energy, Office of Science, Office of Basic Energy Sciences under Contract No. DE-AC02-76SF00515.

%

\end{document}